\documentclass[aps, prd, nofootinbib, amsmath, preprintnumbers, superscriptaddress,secnumarabic, notitlepage,twocolumn]{revtex4-2}
\usepackage[utf8]{inputenc}
\usepackage{verbatim}
\usepackage{graphicx}
\usepackage{xcolor}
\usepackage{amsmath}
\usepackage{amssymb}
\usepackage[a4paper, total={6.5in, 10in}]{geometry}
\usepackage{hyperref}
\usepackage{mathtools}
\usepackage{import}
\usepackage{mathtools}
\usepackage{ulem}
\usepackage{accents}

\usepackage{slashed}

\newcommand{\be}{\begin{equation}}
\newcommand{\ee}{\end{equation}}
\newcommand{\ba}{\begin{array}}
\newcommand{\ea}{\end{array}}
\newcommand{\bea}{\begin{eqnarray}}
\newcommand{\eea}{\end{eqnarray}}

\usepackage[utf8]{inputenc}

\begin{document}
\title{NGC 1068 constraints on neutrino-dark matter scattering}
\author{James M. Cline}
\affiliation{McGill University, Department of Physics, 3600 University Street,
Montr\'eal, QC H3A2T8 Canada}

\author{Matteo Puel}
\affiliation{McGill University, Department of Physics, 3600 University Street,
Montr\'eal, QC H3A2T8 Canada}

\begin{abstract}
The IceCube collaboration has observed the first steady-state point source of high-energy neutrinos, coming from the active galaxy NGC 1068.
If neutrinos interacted strongly enough with dark matter, the emitted
neutrinos would have been impeded by the dense spike of dark matter surrounding the supermassive black hole at the galactic center, which powers the emission.  
We derive a stringent upper limit on the scattering cross section between neutrinos and dark matter based on the observed events and theoretical models of the dark matter spike. 
The bound can be stronger than that obtained by the single IceCube neutrino event from the blazar TXS 0506+056 for some spike models.  
\end{abstract}

\maketitle

\section{Introduction}  Neutrinos and dark matter (DM) are the most weakly coupled particles in the Universe; it is intriguing to imagine that they might interact very weakly with each other.   Much effort has been made to search for signals of $\nu$-DM interactions, through their effects on cosmology
\cite{Mangano:2006mp,Boehm:2013jpa,Bertoni:2014mva,Wilkinson:2014ksa,Mosbech:2020ahp,Hooper:2021rjc}, astrophysics \cite{Fayet:2006sa,Mangano:2006mp,McMullen:2021ikf,Koren:2019wwi,Murase:2019xqi,Carpio:2022sml,Davis:2015rza,Yin:2018yjn}, and direct detection of boosted DM \cite{Zhang:2020nis,Jho:2021rmn,Farzan:2014gza,Das:2021lcr,Ghosh:2021vkt,Bardhan:2022ywd,Lin:2022dbl} and fermionic absorption DM~\cite{Dror:2020czw,Dror:2019onn,Dror:2019dib,PandaX:2022osq,PandaX:2022ood,MAJORANA:2022prn,EXO-200:2022adi,CDEX:2022rxz}.  
Neutrino emission from active galactic nuclei (AGN) can provide a sensitive probe of $\nu$-DM scattering, in particular from the blazar TXS 0506+056,
from which a 290\,TeV neutrino was observed by IceCube \cite{IceCube:2018dnn}.  It was pointed out in Ref.\ \cite{Cline:2022qld} that the ensuing constraints could be
significantly improved by taking into account the dense ``spike'' of DM
that is believed to accrue around the supermassive black hole (SMBH) that powers the AGN engine.

Recently the IceCube collaboration reported observing $79^{+22}_{-20}$ neutrinos from the nearby active galaxy NGC 1068, mostly with energies of
$\sim (1-15)\,$TeV, the first continuously emitting point source of
neutrinos to be discovered \cite{IceCube:2022der}.  Unlike a blazar,
in which the jet points toward Earth, NGC 1068 is a radio galaxy, with the jet pointing $~\sim 90^\circ$ away \cite{Crenshaw:2000pf,jaffe:hal-00002348}, thereby exposing Earth to the equatorial emissions perpendicular to the jet.  The neutrino emission in this case can be dominated by regions closer to the SMBH, where the DM spike plays a more important role.  We thus anticipate that it can further strengthen the constraints on $\nu$-DM scattering, beyond what is possible with a blazar.  In the following, we will show this is indeed the case.

\begin{figure}[t]
\centerline{\includegraphics[scale=0.31]{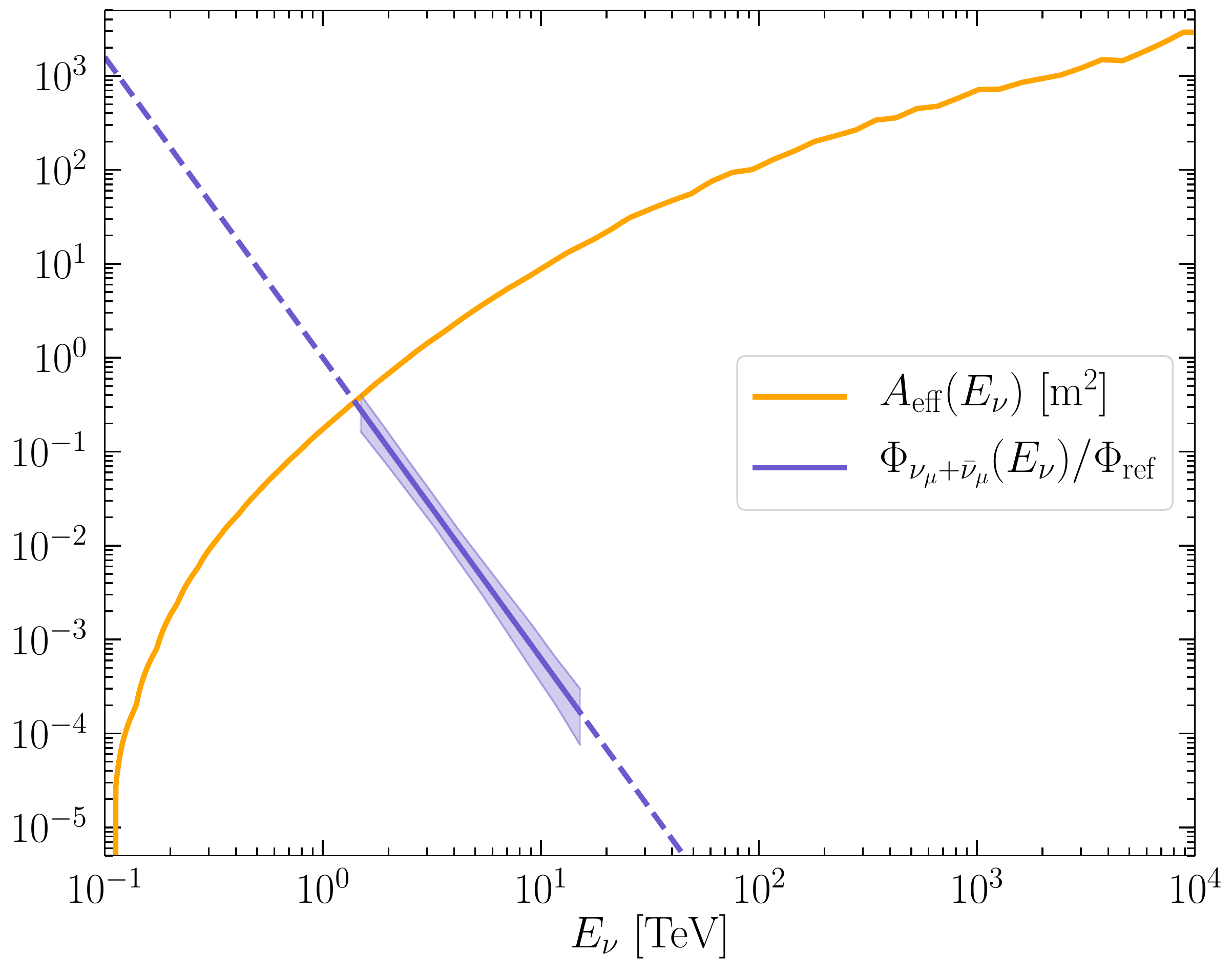}
    }  
    \caption{Normalized neutrino flux $\Phi_{\nu_{\mu} + \bar{\nu}_{\mu}} (E_{\nu}) / \Phi_{\rm ref}$ and IceCube effective area $A_{\rm eff} (E_{\nu})$ from the direction of NGC 1068. The flux, described by Eq.~\eqref{eq:Phi_IceCube}, is shown in slate blue along with its $95\%$ confidence region~\cite{IceCube:2022der}. The orange curve is the effective area, which is taken from Ref.~\cite{icecube-data-ngc1068}.
    } 
    \label{fig:Aeff}
\end{figure}

\section{Neutrino flux}
We use similar methodology to that  in Ref.~\cite{Cline:2022qld}. To set a constraint on the $\nu$-DM scattering cross section using AGNs as neutrino sources one must specify the initial neutrino spectrum and the density of the DM spike.
The IceCube collaboration has 
found that the neutrino flux at Earth is well-described in the energy range $E_{\nu} \in [1.5, 15]$ TeV by an unbroken power-law of the form~\cite{IceCube:2022der}
\be
\label{eq:Phi_IceCube}
    \Phi_{\nu_{\mu} + \bar{\nu}_{\mu}} (E_{\nu}) = \Phi_{\rm ref} \bigg(\frac{E_{\nu}}{E_{\rm ref}} \bigg)^{-\gamma} 
\ee
based on the observation of $\approx 80$ events. Here, $E_{\rm ref} = 1$ TeV, $\Phi_{\rm ref} \approx 5 \times 10^{-11}\,\,\text{TeV}^{-1} \text{cm}^{-2} \text{s}^{-1}$ and $\gamma \approx 3.2$. 
This flux, shown in Fig.~\ref{fig:Aeff}, accounts for the muon neutrino and anti-neutrino contributions only since these are what the IceCube detector can measure; it is expected that all flavors contribute equally to the spectrum, due to neutrino oscillations over cosmic distances, for sources dominated by pion decay~\cite{Bustamante:2015waa,Arguelles:2015dca}. 
Therefore we can view Eq.~\eqref{eq:Phi_IceCube} as the flux where neutrino oscillations are included and the all-flavor flux would be a factor of $3$ higher.

In the following, we will carry out two complementary statistical analyses for setting limits on the $\nu$-DM scattering cross section, based on different reasonable null hypotheses.  In the first approach, we will assume that the initial neutrino flux coincides with the one observed by IceCube, and derive 90\% C.L.\ limits on $\sigma_{\nu\chi}$ by requiring this assumed initial flux not
be attenuated too much.
In the alternative approach, carried out in 
Section~\ref{sec:uncertainties}, we will instead assume the initial flux is that predicted by theoretical models
from the literature.  It will be shown that the two
methodologies give results that are in agreement up to factors of a few.

As a consistency check, we first recomputed the number of expected muon neutrino events observed by IceCube in the case of no $\nu$-DM scattering,
\be
\label{eq:N_pred}
    N_{\rm pred} = t_{\rm obs} \int dE_{\nu}\, \Phi_{\nu} (E_{\nu})\, A_{\rm eff} (E_{\nu})\,,
\ee
where $\Phi_{\nu} \equiv \Phi_{\nu_{\mu} + \bar{\nu}_{\mu}}$ is the flux in Eq.~\eqref{eq:Phi_IceCube}, $t_{\rm obs} \approx 3186$ days \cite{IceCube:2022der} is the observing time, and $A_{\rm eff}$ is the detector's effective area,  shown in Fig.~\ref{fig:Aeff}. 
The relevant events were observed between 2011 and 2020, and data for $A_{\rm eff}$ from the direction of NGC 1068 is provided by IceCube~\cite{icecube-data-ngc1068}.
Integrating Eq.~\eqref{eq:N_pred} over the range $E_{\nu} \in [10^{-1}, 10^4]$ TeV~\cite{IceCube:2022der} gives $N_{\rm pred} \approx 80$ events as expected. Since IceCube considered the flux in Eq.~\eqref{eq:Phi_IceCube} to be reliably determined only within the smaller energy range $E_{\nu} \in [1.5, 15]$ TeV, we find that the corresponding number of predicted muon neutrinos is reduced to $N_{\rm pred} \approx 31 \pm 7$.

\section{Dark matter spike profile}

The second ingredient needed to set a constraint on $\nu$-DM scattering is the density profile of the DM spike. 
The accretion of DM onto the SMBH has been widely studied in the literature, {\it e.g.,} Refs.~\cite{Gondolo:1999ef,Ullio:2001fb,Gorchtein:2010xa,Merritt:2003qk,Gnedin:2003rj,Merritt:2006mt,Shapiro:2022prq,Sadeghian:2013laa,Ferrer:2017xwm}. If the accretion is adiabatic and relativistic effects are neglected, it can be shown that an initially cuspy DM profile of the form
\be
\label{eq:rhoInitial}
    \rho (r) \simeq \rho_s\, \bigg(\frac{r}{r_s} \bigg)^{-\gamma}
\ee
in the region close to the SMBH,
with $r_s$ the scale radius and $\rho_s$ the scale density, evolves into a spike whose density is~\cite{Gondolo:1999ef}
\be
\label{eq:rhoprime}
    \rho' (r) \simeq \rho_R\, \bigg(1 - \frac{4 R_S}{r}\bigg)^3 \bigg(\frac{R_{sp}}{r} \bigg)^{\alpha}\,.
\ee
Here $R_S = 2 G M_{\rm BH}$ is the Schwarzschild radius, $R_{sp}$ is the characteristic size of the spike and $\rho_R \approx \rho'(R_{sp})$.
The slope $\alpha$ of the spike profile in Eq.~\eqref{eq:rhoprime} was found in Ref.~\cite{Gondolo:1999ef} to be related to the slope of the initial profile $\gamma$ in Eq.~\eqref{eq:rhoInitial} by $\alpha = (9 - 2 \gamma) / (4 - \gamma)$. For $0 \leq \gamma \leq 2$, $\alpha$ can vary between $2.25$ to $2.5$ and for a Navarro–Frenk–White (NFW) profile $\gamma = 1$ and $\alpha = 7/3$. 

\subsection{Spike relaxation}

However, the likely presence of a dense stellar component in the inner region of the galaxy can induce gravitational scattering off of DM, causing a softening of the spike profile to a slope of $\alpha = 3/2$, independently of the value of $\gamma$~\cite{Merritt:2003qk,Gnedin:2003rj,Merritt:2006mt,Shapiro:2022prq}.
We consider both the cases of $\alpha = 7/3$ and $\alpha = 3/2$, with $\gamma = 1$.
More precisely, the modification of the slope of the spike due to stellar scattering with DM is relevant only within the influence radius of the SMBH, $r_h$~\cite{Gnedin:2003rj}, which is generally smaller than the size of the spike $R_{sp}$.\footnote{The influence radius is defined as $r_h = G M_{BH}/\sigma_\star^2$,  where
$\sigma_\star$ is the stellar velocity dispersion.
This is the radius within which the
gravitational effects of the SMBH directly affect the motion of the surrounding stars.}
Therefore, for the case of $\alpha = 3/2$, the spike profile in Eq.~\eqref{eq:rhoprime} should be modified by
\be
\label{eq:rhoprime32}
    \rho_{3/2}' (r) \simeq
    \left\{ 
    \begin{aligned} 
    &\rho_N\, \bigg(1 - \frac{4 R_S}{r}\bigg)^3 \bigg(\frac{r_h}{r} \bigg)^{3/2} & r_i \leq r \leq r_h \\
    &\rho_N' \bigg(\frac{R_{sp}}{r} \bigg)^{7/3} & r \geq r_h
    \end{aligned} 
    \right.
\ee
where $r_i = 4 R_S$ is the inner radius of the spike, and the outer profile at $r \geq r_h$ converges to that predicted by Ref.~\cite{Gondolo:1999ef} for $r \geq r_h \gg R_S$, which we rewrite as
\be
\label{eq:rhoprime73}
    \rho_{7/3}' (r) \simeq \rho_N \, \bigg(1 - \frac{4 R_S}{r}\bigg)^3 \bigg(\frac{R_{sp}}{r} \bigg)^{7/3}\,, \qquad r \geq r_i\,.
\ee

\subsection{Effect of DM annihilation}

If DM particles are allowed to annihilate, the spike profile in the innermost region tends to a maximum density $\rho_c = m_\chi/(\langle\sigma_a v\rangle \,t_{\rm BH})$, where $m_\chi$ is the DM mass, $\langle\sigma_a v\rangle$ is an effective annihilation cross section,\footnote{If the dark matter is asymmetric, then the effective $\langle\sigma_a v\rangle=0$, even if the actual $\langle\sigma_a v\rangle$ is large, since in that case annihilations cannot occur within the DM spike.}
and $t_{\rm BH}$ is the age of the SMBH. We take the latter to be $t_{\rm BH} \simeq 10^9$ yrs for NGC 1068, which is well motivated by studies on the mass assembly history of SMBH at high redshift~\cite{Piana_2020}. 
This value is also
similar to that adopted for the blazar TXS 0506+056~\cite{Wang:2021jic,Granelli:2022ysi,Cline:2022qld,Ferrer:2022kei}, which allows a fair comparison between the new limits derived in this paper and the previous ones.  
Considering that the spike profile should merge onto the pre-existing galaxy profile for $r \geq R_{sp}$, which we take as a NFW halo of the form
\be 
\label{eq:rhoNFW}
    \rho_{\rm NFW} (r) = \rho_s\, \bigg(\frac{r}{r_s} \bigg)^{-1} \bigg( 1 + \frac{r}{r_s}\bigg)^{-2}\,,
\ee
then the total DM density can be written as~\cite{Gondolo:1999ef,Ullio:2001fb,Lacroix:2015lxa,Lacroix:2016qpq,Ferrer:2022kei}
\be
\label{eq:rhochi}
    \rho_{\chi}(r) = 
    \left\{ 
    \begin{aligned} 
    &0 \qquad \qquad & r \leq r_i \\
    &\frac{\rho_{\alpha}'(r)\, \rho_c}{\rho_{\alpha}'(r) + \rho_{c}} & r_i \leq r \leq R_{sp} \\
    &\frac{\rho_{\rm NFW} (r)\, \rho_c}{\rho_{\rm NFW} (r) + \rho_{c}} & r \geq R_{sp}
    \end{aligned} 
    \right.
\ee
where $\rho_{\alpha}' (r)$ is given by Eq.~\eqref{eq:rhoprime32} for $\alpha = 3/2$ and by Eq.~\eqref{eq:rhoprime73} for $\alpha = 7/3$.
In Eq.~\eqref{eq:rhochi} we accounted for the possibility that DM annihilation is strong enough to deplete the outer NFW halo as well, in addition to the internal spike.\footnote{We found that the DM profile considered in Ref.~\cite{Ferrer:2022kei} suffers from the lack of the additional DM annihilation in the outer halo for some values of $m_{\chi}$ and $\langle\sigma_a v\rangle$. This problem does not occur for the NGC 1068 case because the outer halo turns out to give a negligible contribution.}

\subsection{Fixing spike parameters}

We have to determine $\rho_N$, $\rho_N'$, $r_h$ and $R_{sp}$ for the spike density $\rho_{\alpha}' (r)$, and $\rho_s$ and $r_s$ for the NFW halo.
Concerning the latter two parameters, they are determined by the virial mass of  NGC 1068, which can be inferred independently of the  DM spike, since the latter cannot appreciably impact the overall halo shape. 
More precisely, N-body simulations in a $\Lambda$CDM universe predict that the DM halo is related to the SMBH mass via~\cite{DiMatteo:2003zx}
\be
\label{eq:DiMatteoMBH-MDM}
   M_{\rm BH} \sim 7 \times 10^7\,M_{\odot} \,\bigg(\frac{M_{\rm DM}}{ 10^{12}\,\,M_{\odot}}\bigg)^{4/3}\,, 
\ee
which is consistent with relations found in Refs.~\cite{Ferrarese:2002ct,Baes:2003rt}. 

Estimates of the NGC 1068 SMBH mass vary.
From the water maser emission line, Ref.~\cite{Greenhill:1996te} estimated a central mass of $M_{\rm BH} \sim 1 \times 10^{7}\,\,M_{\odot}$ within a radius of about $0.65\,\,\text{pc} \simeq 6.5 \times 10^5\,\,R_S$, which we can take as a proxy for the influence radius $r_h$ of the SMBH. Similar values of $M_{\rm BH}$ were found by Refs.~\cite{Greenhill:1997,Hure:2002nu,Lodato:2002cv,Woo:2002un,Panessa:2006sg}, but estimates as large as $M_{\rm BH} \sim (7 - 10) \times 10^7\,\,M_{\odot}$ have been inferred from the polarized broad Balmer and the neutral FeK$\alpha$ emission lines~\cite{Minezaki_2015}. In the following, we consider $M_{\rm BH} = 1 \times 10^7\,\,M_{\odot}$ for which $r_h$ is provided, but we checked that the impact of considering different SMBH masses within the range $\sim 10^7 - 10^8\,\,M_{\odot}$ would impact our results by at most one order of magnitude in the case without DM-annihilation, making them stronger for smaller values of $M_{\rm BH}$. 

From Eq.~\eqref{eq:DiMatteoMBH-MDM} we find that the DM halo mass for NGC 1068 is of order $M_{\rm DM} \simeq 2.3 \times 10^{11}\,\,M_{\odot}$, which is consistent within a factor of $\lesssim 4$ with the predictions from Refs.~\cite{Ferrarese:2002ct,Baes:2003rt,Bullock:1999he,Sabra:2008ygt} and input parameters from Refs.~\cite{Brinks:1997,Paturel:2003zz}.
Taking $M_{\rm DM}$ to be the virial mass for a NFW halo, we can then use Ref.~\cite{Gentile:2007sb} to infer $r_s \simeq 13$ kpc and $\rho_s \simeq 0.35\,\,\text{GeV}/\text{cm}^3$.
The halo parameters we estimated are reasonable in the sense that they are similar to those of the Milky Way galaxy~\cite{Gondolo:1999ef,Ullio:2001fb,Cautun:2019eaf}, which is known to host a SMBH with similar mass to that in NGC 1068~\cite{Event_Horizon_Telescope_Collaboration_2022}.

\begin{table}
\centering
\begin{tabular}{ |c || c | c || c | c |} 
 \hline
 
$\langle\sigma_a v\rangle$ & Model & $\alpha$ & Model & $\alpha$ \\
 \hline
 $0$ & BM1 & $7/3$ & BM$1'$ & $3/2$ \\
 \hline
$0.01$ &  BM2  & $7/3$  & BM$2'$ & $3/2$ \\
 \hline
$3$ &  BM3  & $7/3$ & BM$3'$ & $3/2$ \\
 \hline
\end{tabular}
\caption{Models considered in this paper, distinguished by different values of the effective DM annihilation cross section (in units of $10^{-26}$\,cm$^3$/s) and the spike-profile exponent $\alpha$. They are the same as considered in Ref.~\cite{Cline:2022qld}.  The case $\langle\sigma_a v\rangle\cong 0$ could correspond to asymmetric or freeze-in dark matter. }
\label{tab1}
\end{table}

\begin{figure*}[t]
   \centerline{
   \includegraphics[scale=0.34]{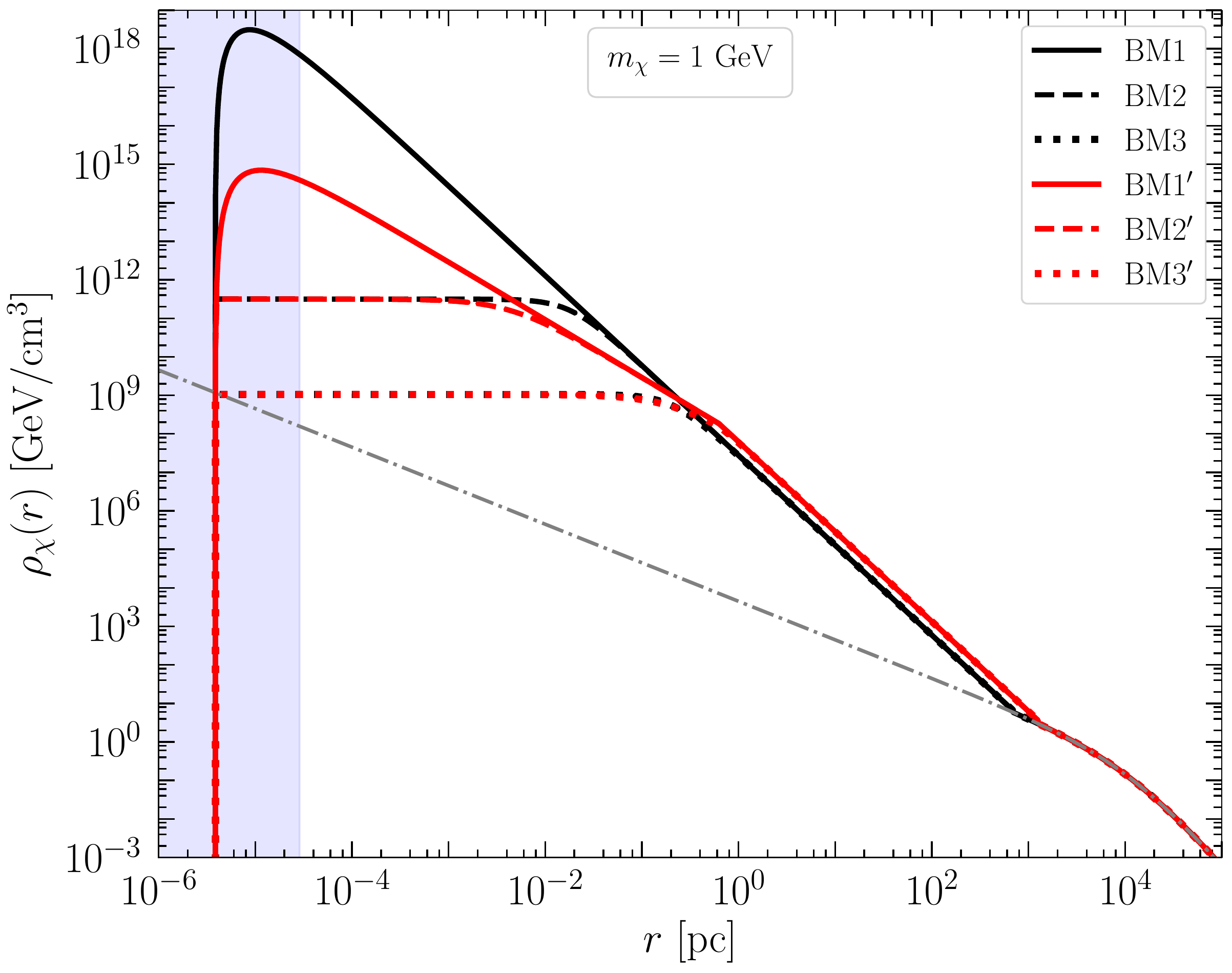}
   \includegraphics[scale=0.34]{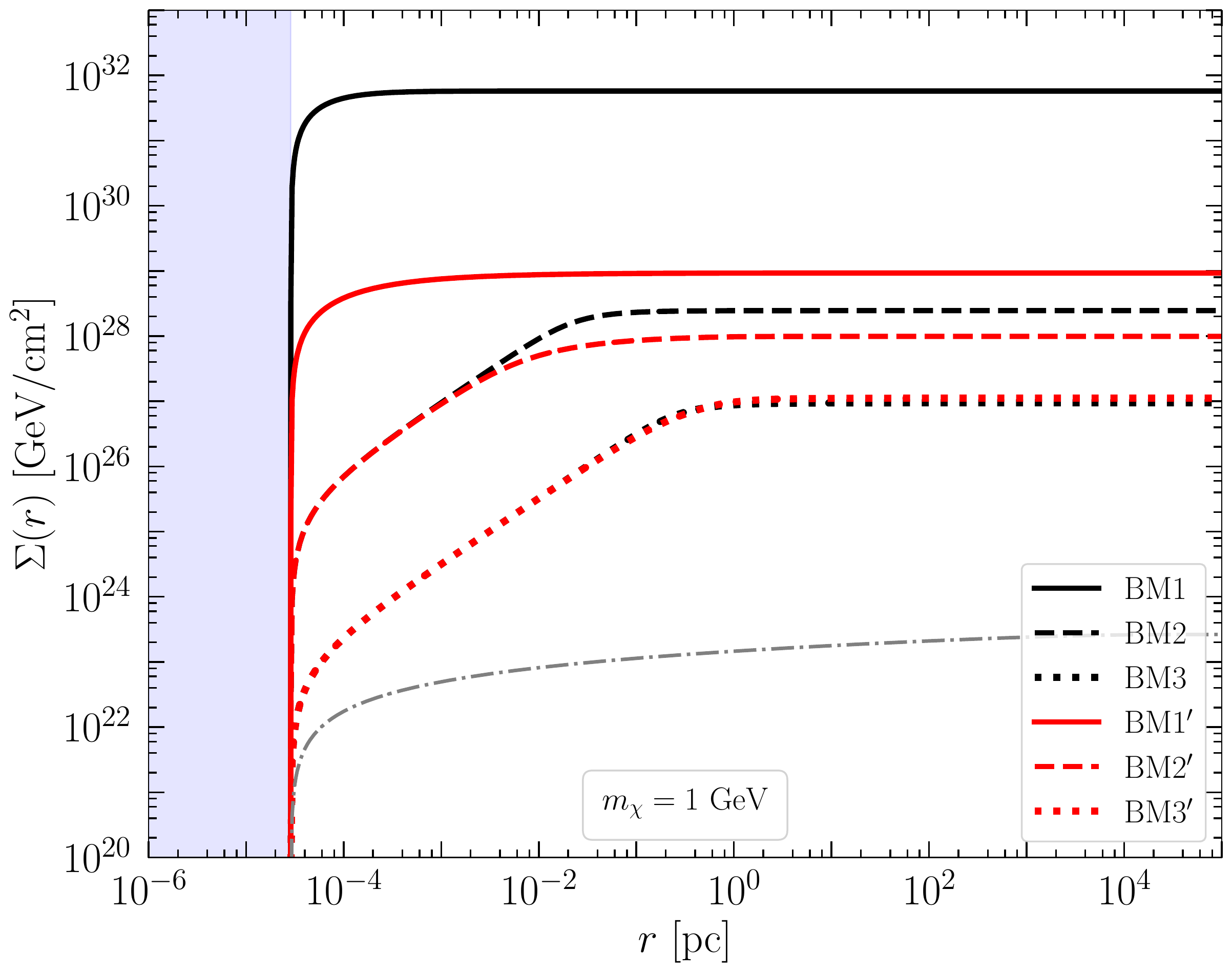}
   }
    \caption{DM density profile $\rho_{\chi} (r)$ for the galaxy NGC 1068 (\textit{left}), given by Eq.~\eqref{eq:rhochi}, and its corresponding accumulated DM column density $\Sigma (r)$ (\textit{right}), given by Eq.~\eqref{eq:Sigmachi}, as a function of the distance from the galactic centre, for the different benchmark models considered in Table~\ref{tab1}.
    The DM mass is fixed to $m_{\chi} = 1$ GeV. The gray dot-dashed curve represents the contribution of the NFW host-halo alone, described by the density in Eq.~\eqref{eq:rhoNFW}. Its contribution to $\Sigma (r)$ is negligible compared to that of the spike. The blue shaded region delimits the region within the galaxy where neutrinos are not likely to be emitted. Its right edge corresponds to the value of $R_{\rm em}$ used in Eq.~\eqref{eq:Sigmachi}. 
    }  
    \label{fig:rhoSigma_vs_r}
\end{figure*}

The normalization $\rho_N$ of the profiles in Eqs.~\eqref{eq:rhoprime32} and~\eqref{eq:rhoprime73} depends only upon the combination $\mathcal{N} \equiv \rho_N\, r_{h}^{3/2}$ or $\mathcal{N} \equiv \rho_N\, R_{sp}^{7/3}$, respectively, which can be determined by the mass of the SMBH~\cite{Ullio:2001fb,Gorchtein:2010xa,Lacroix:2015lxa,Lacroix:2016qpq,Cermeno:2022rni}
\be
\label{eq:M_BH}
    M_{\rm BH} \approx 4 \pi \int_{r_i}^{r_h} dr\,r^2\,\rho_{\chi} (r)\,,
\ee
where $r_h$ is the radius of influence of the black hole and $\rho_{\chi} (r)$ is given by Eq.~\eqref{eq:rhochi}. 
The integral in Eq.~\eqref{eq:M_BH} is dominated by the contribution from $r \gg r_i$, in which regime $\rho_{\chi} (r) \simeq \rho_{\alpha}' (r)$ since $\rho_c \gg \rho_{\alpha}'(r)$. 
Therefore, we find that $\mathcal{N}$ obtained from Eq.~\eqref{eq:M_BH} is approximately given by
\be
\label{eq:Nnorm}
    \mathcal{N} \simeq \frac{M_{\rm BH}}{4 \pi\,[f_{\alpha}(r_h) - f_{\alpha}(r_i)]}\,,
\ee
where 
\be 
    f_{\alpha}(r) \equiv r^{-\alpha} \bigg(\frac{r^3}{3 - \alpha} + \frac{12 R_S\, r^2}{\alpha - 2} - \frac{48 R_S^2\, r}{\alpha - 1} + \frac{64 R_S^3}{\alpha} \bigg)\,.
\ee
For the $\alpha = 3/2$ case, the normalization density $\rho_{N}'$ in Eq.~\eqref{eq:rhoprime32} is determined by requiring the profile $\rho_{3/2}' (r)$ to match at $r = r_h$, giving $\rho_N' \simeq \rho_N\, (r_h / R_{sp})^{7/3}$  $\simeq \mathcal{N}\,r_h^{5/6} / R_{sp}^{7/3}$.
On the other hand, the value of $R_{sp}$ can be obtained from the matching condition between the spike and the outer halo, namely $\rho_N^{\prime},\, \rho_N \simeq \rho_s (R_{sp} / r_s)^{-\gamma}$ for $\alpha = 3/2,\, 7/3$, respectively, since $R_S \ll r_h < R_{sp} \ll r_s$. This translates into 
\be
    R_{sp} \simeq 
     \left\{ 
    \begin{aligned} 
    &\bigg(\frac{\mathcal{N}}{\rho_s r_s} \bigg)^{3/4}\,r_h^{5/8}\,, \qquad & \alpha = 3/2 \\
    &\bigg(\frac{\mathcal{N}}{\rho_s r_s} \bigg)^{3/4}\,, & \alpha = 7/3
    \end{aligned} 
    \right.\,,
\ee
where $\mathcal{N}$ is given by Eq.~\eqref{eq:Nnorm}. We find that $R_{sp} \sim 0.7\,\,\text{kpc} \gg r_h$ for the parameter values used in this paper.
The DM profile of NGC 1068, described by Eq.~\eqref{eq:rhochi}, is shown in the left panel of Fig.~\ref{fig:rhoSigma_vs_r} for different choices of the effective DM annihilation cross section $\langle\sigma_a v\rangle$ and the spike-profile exponent $\alpha$, according to the benchmark models considered in Ref.~\cite{Cline:2022qld} and summarized in Table~\ref{tab1}.

\subsection{DM column density}

An important quantity for $\nu$-DM scattering is the DM column density, which is the projection of the 3D density of DM along the line-of-sight, integrating over the
distance that neutrinos travel during their journey to Earth. The accumulated column density is defined by
\be
\label{eq:Sigmachi}
    \Sigma (r) = \int_{R_{\rm em}}^{r} dr'\, \rho_{\rm DM} (r') \approx \int_{R_{\rm em}}^{r} dr'\, \rho_{\rm \chi} (r')\,,
\ee
where $R_{\rm em}$ is the distance from the SMBH to the position in the radio galaxy where neutrinos are first likely to be produced. 
Neutrinos with energy below $100$ TeV are believed to be generated through hadronic interactions occurring in the AGN corona~\cite{Katz:1976,Bisnovatyi-Kogan:1977,Eichler:1979,Pozdnyakov:1983,Galeev:1979,Begelman:1990,Stecker:1991,Kalashev:2014vya,Inoue:2019yfs,Inoue:2019fil,Murase:2019vdl,Kheirandish:2021wkm,Murase:2015xka}, whose radius has been estimated to be a few tens of the Schwarzschild radius $R_S$~\cite{Murase:2019vdl,Inoue:2019yfs,Inoue:2019fil,Inoue:2018kbv,Gallimore:2004wk,Murase:2022dog}. We take $R_{\rm em} \simeq 30\, R_S$ as in Ref.~\cite{Murase:2019vdl}, whose neutrino spectrum matches well the IceCube observation from NGC 1068, and supported by a model-independent study in Ref.~\cite{Murase:2022dog}.
In the second step of Eq.~\eqref{eq:Sigmachi}, we neglected the cosmological and Milky-Way DM contributions to $\Sigma (r)$ (see for instance Ref.~\cite{Choi:2019ixb}) because they are orders of magnitude smaller than that given by the DM spike profile~\cite{Cline:2022qld,Ferrer:2022kei}.

The right panel of Fig.~\ref{fig:rhoSigma_vs_r} shows the accumulated DM column density as a function of the distance from the central SMBH, for the different benchmark models of Table~\ref{tab1}.
Given that $\Sigma (r)$ stays constant for $r \gtrsim \mathcal{O}(100\,\,\text{pc})$, similarly to that found in Ref.~\cite{Wang:2021jic,Granelli:2022ysi} for the blazar TXS 0506+056, we can take the DM column density at the Earth location, roughly distant $D_L\simeq 14.4$ Mpc from NGC 1068~\cite{Tully:1988,Bland-Hawthorn:1997}, to be approximately
$\Sigma(D_L) \cong \Sigma (r \simeq 100\,\,\text{pc}) \equiv \Sigma_\chi$.

In contrast to Ref.~\cite{Ferrer:2022kei}, we find that the inclusion of the halo of the host galaxy (which becomes important for $r \gtrsim R_{sp}$) gives a negligible contribution to $\Sigma_{\chi}$, as also suggested by Ref.~\cite{Granelli:2022ysi}. This can be explained by the fact that the normalization of the outer halo $\rho_s$ cannot be inferred through the 
the SMBH mass, given by Eq.~\eqref{eq:M_BH}, since the latter can constrain at most the DM spike normalization density $\rho_N$. Instead, $\rho_s$ should be chosen so that the total DM mass of the host halo matches reasonable values as found by observations and cosmological simulations, since the spike makes a negligible contribution to the total halo mass. 

We note that a more accurate description of the adiabatic accretion of DM onto SMBH would require the inclusion of relativistic effects. Ref.~\cite{Sadeghian:2013laa} found that not only the inner radius of the DM spike reduces to $r_i = 2 R_S$ from its nonrelativistic value of $4 R_S$, but also the spike reaches significantly higher densities. For a rotating SMBH, the density enhancement is even more important~\cite{Ferrer:2017xwm}. However, such relativistic effects become irrelevant for distances $r \gtrsim 20\,R_S$~\cite{Sadeghian:2013laa}, which is where neutrinos are likely to be produced in NGC 1068.

\begin{figure*}[t]
\centerline{\includegraphics[scale=0.34]{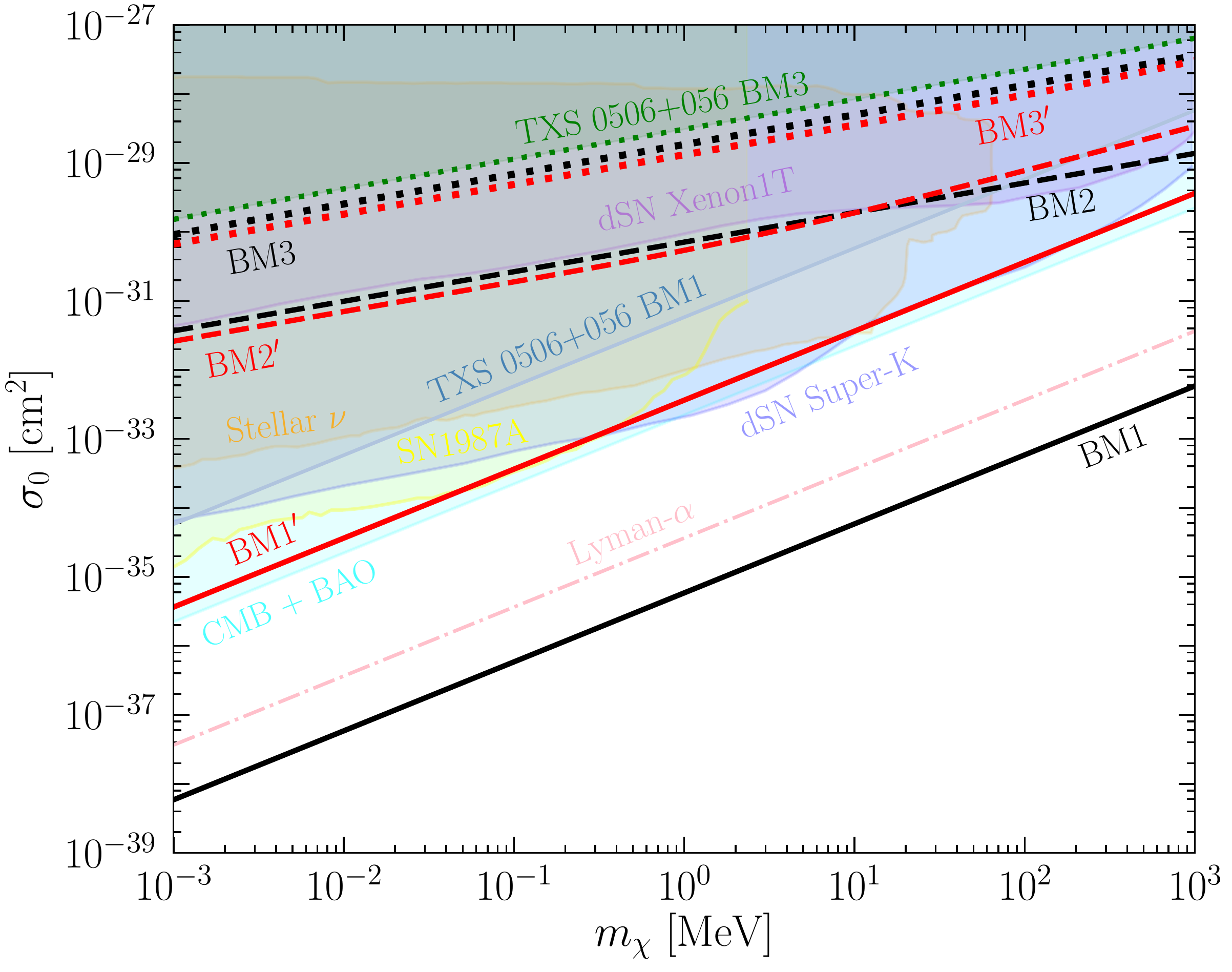}
   \includegraphics[scale=0.34]{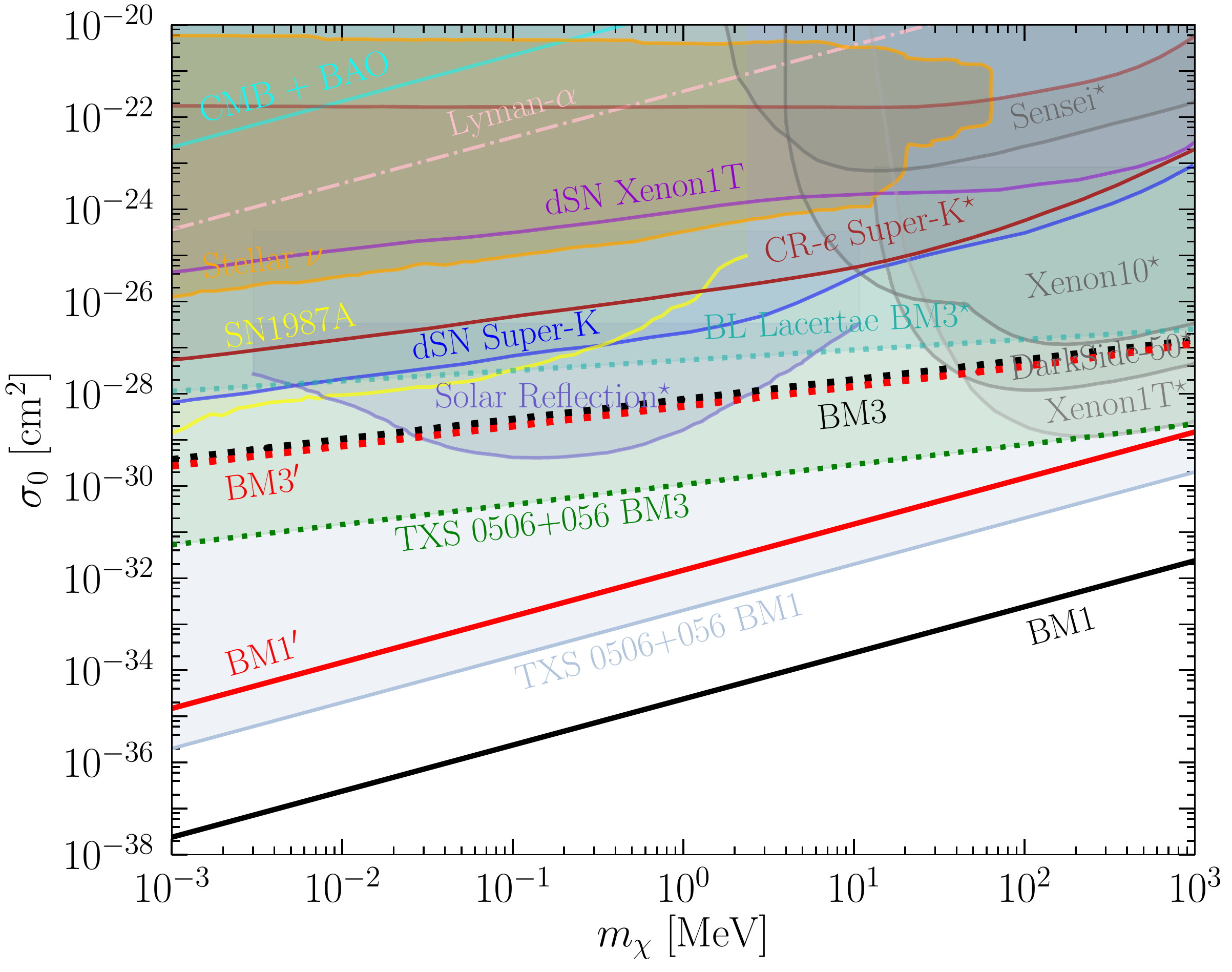}}
    \caption{\textit{Left:} $90\%$ C.L. upper limits on the $\nu$-DM
    scattering cross section, assumed to be energy-independent, for
    the six benchmark DM spike models of Table~\ref{tab1}, compared to previous constraints. The latter are:
    (cyan) CMB and baryon acoustic oscillations~\cite{Mosbech:2020ahp}; (pink) Lyman-$\alpha$
    preferred model~\cite{Hooper:2021rjc}; (dark violet, blue) diffuse
    supernova neutrinos~\cite{Ghosh:2021vkt}; (orange) stellar
    neutrinos~\cite{Jho:2021rmn}; (yellow) supernova
    SN1987A~\cite{Lin:2022dbl}; (green, light steel blue) least and most restrictive bounds, namely for BM3 and BM1 respectively, from TXS 0506+056
    \cite{Cline:2022qld}.
    \textit{Right:} Same, but assuming linear energy-dependent $\nu$-DM and $e$-DM scattering cross-sections. All the constraints are rescaled to the energy $E_0 = 10$ TeV according to the relation $\sigma_{\nu \chi} = \sigma_0 (E_{\nu}/E_0)$. Only the least (BM3, BM3$^{\prime}$) and most restrictive (BM1, BM1$^{\prime}$) new limits are shown.
    The $\nu$-DM scattering bounds are the same as those in the left panel, while for $e$-DM scattering they are labelled with $\star$ and are: (slate blue) solar reflection~\cite{An:2017ojc}, (brown) Super-K for DM boosted by cosmic-ray electrons, (turquoise) blazar BL Lacertae for BM3 model~\cite{Granelli:2022ysi}, (gray) direct detection for light DM interacting with electrons~\cite{SENSEI:2020dpa,Essig:2017kqs,DarkSide-50:2022hin,XENON:2019gfn}.
    }  
    \label{fig:limit1}
\end{figure*}

\section{Constraints on neutrino-dark matter scattering}
If neutrinos scatter with DM along their journey to the detector, the attenuation of the neutrino flux can be described by a form of the Boltzmann equation known as the cascade equation~\cite{Arguelles:2017atb,Vincent:2017svp,Cline:2022qld}
\be
    \frac{d\Phi}{d\tau}(E_\nu) = -\sigma_{\nu\chi}\Phi + \int_{E_\nu}^\infty dE'_\nu\,
        \frac{d\sigma_{\nu\chi}}{dE_\nu}({\scriptstyle E'_\nu\to E_\nu})\, \Phi(E'_\nu)\,,
        \label{cascade}
\ee  
where $\tau = \Sigma (r)/m_\chi$ is the accumulated
column density, defined in Eq.~\eqref{eq:Sigmachi}, over the DM mass and $E_{\nu}$ and $E_{\nu}'$ are the energies of the outgoing and incoming neutrino, respectively. 
The first term of the cascade equation describes the flux depletion of neutrinos with energy $E_{\nu}$ due to their interaction with the surrounding DM, whereas the second term corresponds to the effect of energy redistribution of the neutrinos from high to low energy. 

The cascade equation is designed to predict the change in the energy spectrum due to downscattering, for high-energy beams that are scattered predominantly in the forward direction.  If $d\sigma_{\nu\chi}/dE_{\nu}$ is nonzero, then it can be seen that $\int_{0}^{\infty} dE_\nu d\Phi/d\tau = 0$, indicating that the total number of particles in the beam is conserved, and only their spectrum changes. 
On the other hand, if one sets $d\sigma_{\nu\chi}/dE_{\nu} = 0$, the equation describes the inelastic process of neutrino absorption in which there is no final-state energy $E_{\nu}$. 
For elastic scattering with constant cross section, the neutrino does not lose energy in the process (only its direction changes), translating into $d\sigma_{\nu \chi} / dE_{\nu} \propto \delta (E_{\nu}' - E_{\nu})$ and resulting in a conserved flux according to eq.~\eqref{cascade}.

As will be shown in 
Section \ref{sec:Zp_model}, another situation where the cross section is approximately constant is the
high-energy regime of a model with a mediator of mass 
$m_{Z'}\ll\sqrt{E_\nu m_\chi}$.  Even though $d\sigma/dE_\nu$ is not strictly zero in that case,
we showed in Ref.\ \cite{Cline:2022qld}
that to a good approximation, it can be neglected for such high energies, and we will use this approximation where appropriate below.
Including or neglecting the second term in the cascade equation therefore corresponds to different microscopic origins of the scattering,
and the results should be interpreted accordingly when comparing to specific particle physics models.

We consider two different cases that describe important regimes for the cross section in particular particle physics models~\cite{Cline:2022qld}:
\begin{enumerate}
    \item[(a)] $\sigma_{\nu \chi} \simeq \sigma_0 = \text{const}$;
    \item[(b)] $\sigma_{\nu \chi} \simeq \sigma_0\,(E_{\nu} / E_0)$, with $E_0 = \text{const}$.
\end{enumerate}
For the former case, Eq.~\eqref{cascade} leads to an exponential attenuation of the neutrino flux according to $\Phi (E_{\nu}) \sim \Phi_{\nu} (E_{\nu})\, \exp{(- \sigma_0\,\Sigma_{\chi} / m_{\chi})}$, where $\Phi_{\nu} (E_{\nu})$ is the initial flux at the source location, given by Eq.~\eqref{eq:Phi_IceCube}, and $\Phi (E_{\nu})$ is the final flux at the detector location. 

To set limits on the scattering cross section, we require the number of neutrinos reaching the detector at Earth to be no less than what IceCube observed. In particular, the $90\%$ Confidence Level (C.L.) upper limit on $\sigma_{0}$ follows from demanding that the number of events in the energy range $E_{\nu} \in [1.5, 15]$ TeV exceed $\simeq 22.2$; this is
the $90\%$ C.L. lower limit on the number $31 \pm 7$ of observed IceCube events, 
assuming a Gaussian distribution.
For energy-independent scattering and initial flux given by Eq.~\eqref{eq:Phi_IceCube}, this corresponds to a flux attenuation of $\sim 30\%$, resulting in the bound~\footnote{For the IceCube observation of a single neutrino from the blazar TXS 0506+056, the $90\%$ C.L. lower limit on the Poisson-distributed number of observed events is $\simeq 0.1$~\cite{Feldman:1997qc}, as used in Ref.~\cite{Cline:2022qld}. This corresponds to a $90\%$ absorption of the initial flux if the scattering cross section is energy-independent, as previously considered in the literature~\cite{Choi:2019ixb,Ferrer:2022kei,Cline:2022qld}.}

\be
\label{eq:sigma0const}
    \sigma_0 < 0.34 
    \, {m_{\chi} \over\Sigma_{\chi}}\,.
\ee
The left panel of Fig.~\ref{fig:limit1} shows the resulting constraints for the different benchmark models of Table~\ref{tab1}.


The new bounds for the four annihilating DM models (BM2, BM2$^{\prime}$, BM3 and BM3$^{\prime}$) are a factor $2$-$3$ stronger than those inferred from the blazar TXS 0506+056 by Ref.~\cite{Cline:2022qld}, as can be seen by comparing the black dotted curve with the dotted green one in Fig.~\ref{fig:limit1}.  On the other hand, the constraint from
TXS 0506+056 on $\sigma_0 \Sigma_\chi/m_\chi$ was $\sim 50$ times stronger than the analogous one
derived here, Eq.\ \eqref{eq:sigma0const}.  The apparent
discrepancy arises because 
$\Sigma_{\chi}$ for NGC 1068 is much greater than that for TXS 0506+056. Indeed,
the total DM density profile $\rho_{\chi} (r)$ for TXS 0506+056 peaks at larger distances from the galactic centre than that for NGC 1068, with a smaller spike amplitude, given that the mass of the SMBH within the former galaxy is larger than the one in NGC 1068, and so are the corresponding Schwarzschild radii. 
In other words, the smaller the black hole mass, the larger is the effect of the DM spike, which would peak at a distance closer to the galactic centre compared to a heavier black hole.
This effect is even more pronounced in the the nonannihilating DM models BM1 and BM1$^{\prime}$, which
give stronger limits than those obtained in Ref.~\cite{Cline:2022qld} for TXS 0506+056,
thanks also to the smaller value of $R_{\rm em}$ for the radio galaxy than that for the blazar. In
particular the limit for the model BM1 surpasses any existing bounds on $\sigma_{\nu \chi}$ in the literature by several orders of magnitude.

For the linear energy-dependent $\sigma_{\nu \chi} = \sigma_0(E_\nu/E_0)$ case, the cascade equation~\eqref{cascade} can be solved by discretizing it in equal logarithmic energy intervals and using the methods outlined in Refs.~\cite{Cline:2022qld,Vincent:2017svp}. The $90\%$ C.L. limit on $\sigma_0$ is derived in the same way as described for the energy-independent case.
Taking $E_0 = 10$ TeV, which lies within the range of validity of the IceCube flux in Eq.~\eqref{eq:Phi_IceCube}, we find that 
\be
\label{eq:sigma0lin}
\sigma_0 < 1.25 
\, \frac{m_{\chi}}{\Sigma_{\chi}}\,,
\ee
which is roughly a factor of four times weaker than the energy-independent scattering case considered above, given by Eq.~\eqref{eq:sigma0const}. The corresponding limits on $\sigma_0$ for the different benchmark models are shown on the right panel of Fig.~\ref{fig:limit1}, where the strongest constraints on the $\nu$-DM scattering existing in the literature are included, after being rescaled to the common energy scale $E_0$ using the relation $\sigma_{\nu \chi} = \sigma_0\,(E_{\nu}/ E_0)$. 

Bounds on electron-DM scattering are also included in the right panel of Fig.\ \ref{fig:limit1} (labelled with $\star$), since it is natural for DM to interact with neutrinos and charged leptons with equal strength, due to the SU$(2)_L$ gauge symmetry of the standard model. We refer the reader to Ref.~\cite{Cline:2022qld} for the description of the existing limits.
We do not include the recent bounds on $\sigma_{e\chi}$ inferred by Ref.~\cite{Bhowmick:2022zkj} because they are model-dependent.
Our new limits on $\sigma_0$, shown in the right panel for the case of $\sigma_{\nu\chi} \propto E_{\nu}$, are generally weaker than the constraints implied by the $290$ TeV neutrino event from  TXS 0506+056, except for the most optimistic BM1 spike model~\footnote{It was recently argued that the $290$ TeV IceCube neutrino from the blazar TXS 0506+056 might have  originated from a region closer to the central black hole, such as the corona, rather than coming from the jet \cite{Halzen}. This is based on the observation by the MASTER Collaboration of a rapid change in the optical luminosity of the blazar between one minute and two hours after the IceCube detection~\cite{Lipunov:2020ptp}, which raises the question whether such fast variability could occur at very large distances from the central engine, pertaining to the jet. If this was indeed true, the bounds on $\sigma_0$ derived in Ref.~\cite{Cline:2022qld} at energy $E_0 = 290$ TeV would become stronger by about a factor of 50 (5) for the BM$1$ (BM$1'$) model, assuming the initial neutrino flux used there can still be valid. We thank Francis Halzen for brining this to our attention.}.

\begin{table*}[t]
\centering
\begin{tabular}{|c | c | c | c |} 
 \hline
 
Color in Fig.~\ref{fig:initial_fluxes} 
& Ref. of $\Phi_{\nu}$ (color curve) & $\sigma_{\nu \chi} \sim \text{const}$ [$m_{\chi} / \Sigma_{\chi}$] & $\sigma_{\nu \chi} \sim E_{\nu}$ [$m_{\chi} / \Sigma_{\chi}$] \\
 \hline
 \hline
 slate blue (IceCube) & Fig. 4 of \cite{IceCube:2022der} (slate blue) & $\sigma_0 < 0.34$ & $\sigma_0 < 1.25$ \\
 \hline
 brown & Fig. 2 of \cite{Inoue:2019yfs} (blue, upper) & $\sigma_0 < 0.27$ & $\sigma_0 < 1.33$ \\
 \hline
 lime & Fig. 1 of \cite{Kheirandish:2021wkm} (red) & $\sigma_0 < 0.36$ & $\sigma_0 < 1.13$ \\
 \hline
 hot pink & Fig. 4 of \cite{Kheirandish:2021wkm} (deep pink, upper) & $\sigma_0 < 1.52$ & $\sigma_0 < 5.32$ \\
 \hline
 medium orchid & Fig. 12 of \cite{Kheirandish:2021wkm} (green, thin) & $\sigma_0 < 0.11$ & $\sigma_0 < 0.34$ \\
 \hline
 deep sky blue & Fig. 5 of \cite{Eichmann:2022lxh} (black-green) & $\sigma_0 < 0.44$ & $\sigma_0 < 1.67$ \\
 \hline
 crimson & Fig. 3 left of \cite{Murase:2022dog} (black) & $\sigma_0 < 0.29$ & $\sigma_0 < 1.09$ \\
 \hline
 gold & Fig. 3 right of \cite{Murase:2022dog} (black) & $\sigma_0 < 0.17$ & $\sigma_0 < 0.64$ \\
 \hline
 green & Fig. 6 of \cite{Inoue:2022yak} (green, dashed) & $\sigma_0 < 0.30$ & $\sigma_0 < 1.06$ \\
 \hline
 orange & Fig. 6 of \cite{Inoue:2022yak} (orange, solid) & $\sigma_0 < 0.03$ & $\sigma_0 < 0.10$ \\
 \hline
\end{tabular}
\caption{$90\%$ C.L. upper bounds on $\sigma_0$ (in units of $m_{\chi} / \Sigma_{\chi}$) for different initial neutrino fluxes $\Phi_{\nu}$, both for energy-independent and energy-dependent scattering cross sections. The first column contains the color of the curve used in Fig.~\ref{fig:initial_fluxes}, the second column displays the reference and figure where each input flux is taken and what color is used to show the curve there. The third and fourth columns contain our derived bounds on $\sigma_0$. The first row corresponds to the bounds in Eqs.~\eqref{eq:sigma0const} and~\eqref{eq:sigma0lin} using the IceCube observed flux in Eq.~\eqref{eq:Phi_IceCube} as the input spectrum.}
\label{tab:diff_fluxes}
\end{table*}

\begin{figure*}[t]
\centerline{\includegraphics[scale=0.34]{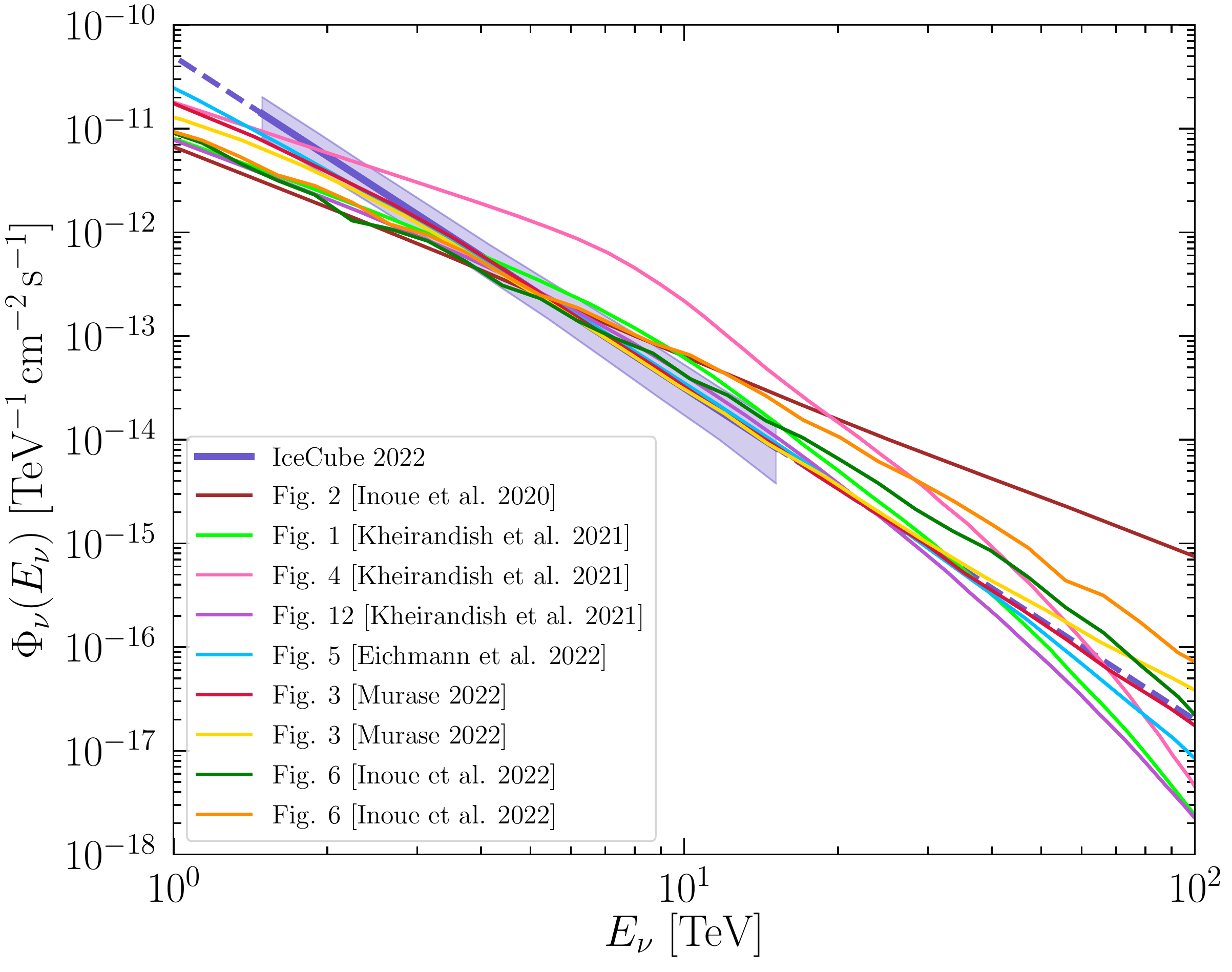}
\includegraphics[scale=0.34]{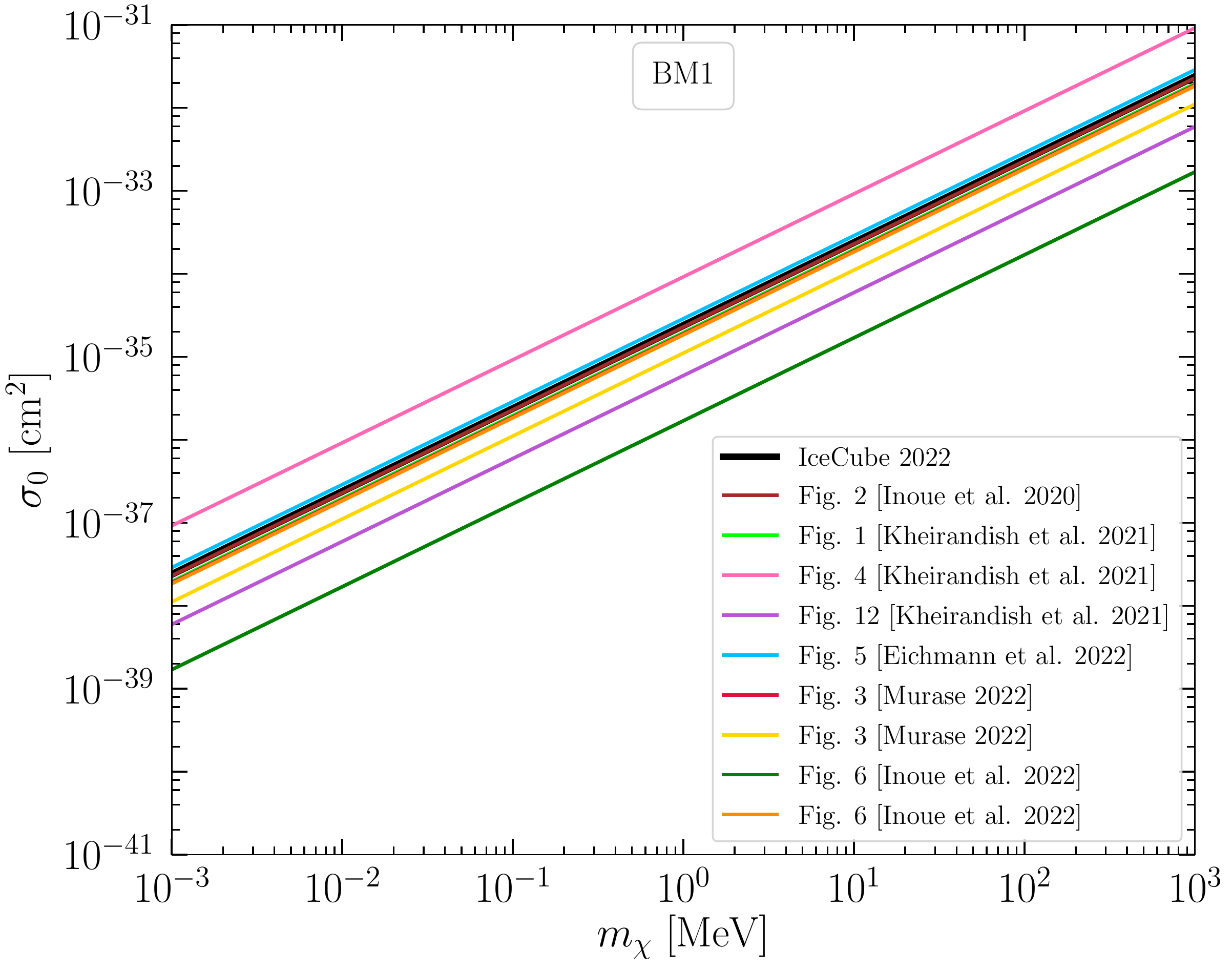}}
    \caption{\textit{Left:} The initial neutrino fluxes $\Phi_{\nu}$ used in this paper and summarized in Table~\ref{tab:diff_fluxes}. 
    The corresponding models used to explain the observational data are, in order of appearance: (slate blue) corona $p p$ scenario with gyrofactor $\eta_g = 30$, which is the mean free path of a particle in units of the gyroradius~\cite{Inoue:2019yfs}, (lime) stochastic scenario with high cosmic ray (CR) pressure and with x-ray luminosity $L_{X} = 10^{43}$ erg/s~\cite{Kheirandish:2021wkm}, (hot pink) same with $L_{X} = 10^{43.8}$ erg/s~\cite{Kheirandish:2021wkm}, (medium orchid) magnetic reconnection fast acceleration scenario with injected CR power-law exponent $s = 1$, acceleration efficiency $\eta_{\rm acc} = 300$ and maximum proton energy $E_p^{\rm rec} = 0.1$ PeV~\cite{Kheirandish:2021wkm}, (deep sky blue) corona plus starburst model~\cite{Eichmann:2022lxh}, (crimson) minimal $p p$ scenario~\cite{Murase:2022dog}, (gold) minimal $p \gamma$ scenario~\cite{Murase:2022dog}, (green) wind plus torus model with magnetic field strength $B = 1130$ G and gyrofactor $\eta_g = 4$~\cite{Inoue:2022yak}, (orange) same with $B = 510$ G and $\eta_g = 1$.
    The slate blue curve is the IceCube observed flux in Eq.~\eqref{eq:Phi_IceCube} as shown in Fig.~\ref{fig:Aeff}.
    \textit{Right:} $90\%$ C.L. limits on the cross section $\sigma_0$ at energy $E_{\nu} = 10$ TeV in the linear energy-dependent scattering scenario, for the spike model BM$1$ and for different choices of the initial flux $\Phi_{\nu}$. The black curve is the result shown in the right panel of Fig.~\ref{fig:limit1} using the IceCube observed flux in Eq.~\eqref{eq:Phi_IceCube} as the input spectrum.
    The rest of the color coding is the same in both panels.
    }
    \label{fig:initial_fluxes}
\end{figure*}

\section{Astrophysical uncertainties}
\label{sec:uncertainties}

The results derived in the previous sections and summarized in Fig.~\ref{fig:limit1} are subject to astrophysical uncertainties, which we elaborate on in this section.
So far, we considered variations in  the shape of the DM spike profile, by allowing for a range of possible values of the profile exponent $\alpha$ and DM annihilation cross section $\langle \sigma_a v \rangle$.  Another parameter is the size of the emission region, characterized by the radial distance $R_{\rm em}$ from the central black hole beyond which
neutrinos are likely to be produced.  A third is the initial neutrino flux $\Phi_{\nu}$.  In the foregoing analysis, we kept the latter two quantities fixed, namely to $R_{\rm em} \simeq 30\,R_S$ and $\Phi_{\nu} = \Phi_{\nu_{\mu} + \bar{\mu}_{\nu}}$, as given by Eq.~\eqref{eq:Phi_IceCube}.

A largely model-independent study on the connection between neutrinos and gamma rays in NGC 1068, based on the recent IceCube $\approx 80$ events, found that neutrinos in the TeV range are most likely to be emitted from regions within $\sim 30 - 100\,R_S$ from the galactic center~\cite{Murase:2022dog}. Dimensional analysis arguments on neutrino production in AGN obscured cores reach similar conclusions~\cite{Halzen}.
Allowing for $R_{\rm em} \simeq 100\, R_S$, the value of $\Sigma (r)$ implied by Eq.~\eqref{eq:Sigmachi} would decrease by a factor of $\sim 4$ ($2$) for the non-annihilation DM model BM1 (BM1$^{\prime}$), weakening the corresponding constraints on $\sigma_0$ by the same factor.

Concerning the initial neutrino flux, we recomputed the limits by considering dozens of predicted neutrino spectra for NGC 1068 found in the literature.  These vary in terms of source modeling, production mechanisms for both photons and neutrinos, and model parameters~\cite{Murase:2019vdl,Inoue:2019yfs,Kheirandish:2021wkm,Anchordoqui:2021vms,Eichmann:2022lxh,Inoue:2022yak,Murase:2022dog,Yoast-Hull:2013qfa,Lamastra:2016axo,MAGIC:2019fvw}. 
Among them, only the nine examples shown in the left panel of Fig.~\ref{fig:initial_fluxes} give at least as many expected muon neutrinos as observed by IceCube within the energy range $E_{\nu} \in [1.5, 15]$ TeV, according to Eq.~\eqref{eq:N_pred} in the absence of $\nu$-DM scattering. 
Carrying out the same analysis as in the previous section, we derived the $90\%$ C.L. upper limits on $\sigma_0$ for all of these initial neutrino spectra, for both the energy-independent and linear energy-dependent scattering cases.

The results are summarized in Table~\ref{tab:diff_fluxes} and shown in the right panel of Fig.~\ref{fig:initial_fluxes} for the 
example of the BM1 spike model. Although there is a factor of $\sim 10 - 15$ ($50 - 60$) scatter in the upper limits of $\sigma_0$ for the energy-dependent (energy-independent) scattering cross section limit inferred from the nine fluxes, only one of them provides a weaker limit (by a factor of $4 - 5$) than we obtained using Eq.~\eqref{eq:Phi_IceCube} as the input flux. 
The associated spectrum, taken from Fig.\ 4 of Ref.~\cite{Kheirandish:2021wkm} (hot pink curve in Fig.~\ref{fig:initial_fluxes}), was derived within the stochastic acceleration scenario with high cosmic ray pressure for the largest x-ray luminosity $L_{X} \simeq 10^{43.8}$ erg/s allowed by observations~\cite{Marinucci:2015fqo}; hence it could be considered to be an outlier. We refer the reader to Ref.~\cite{Kheirandish:2021wkm} for model details.

Since all but one of the available initial neutrino spectra give stronger constraints on $\sigma_0$ than we obtained using the IceCube observed flux in Eq.~\eqref{eq:Phi_IceCube}, the latter can be viewed as not only the most model-independent option, but it also gives a conservative choice for setting limits on $\nu$-DM interactions~\footnote{Most, if not all, of the initial neutrino fluxes we tested were tuned to reproduce both the NGC 1068 electromagnetic data~\cite{Prieto:2010,Fermi-LAT:2017sxy,Fermi-LAT:2019yla,MAGIC:2019fvw,Marinucci:2015fqo} and the IceCube neutrino data~\cite{IceCube:2019cia,IceCube:2022der}. It is possible that by relaxing the requirement that the neutrino spectrum emitted by the source should be consistent with IceCube observations, the flux normalization may be higher than that assumed so far in the literature, giving more room for nonstandard neutrino interactions, such as $\nu$-DM scattering, to match IceCube data and potentially weakening the constraints derived in this paper. We leave this question for future investigation.}.

\begin{figure*}[t]
\centerline{\includegraphics[scale=0.34]{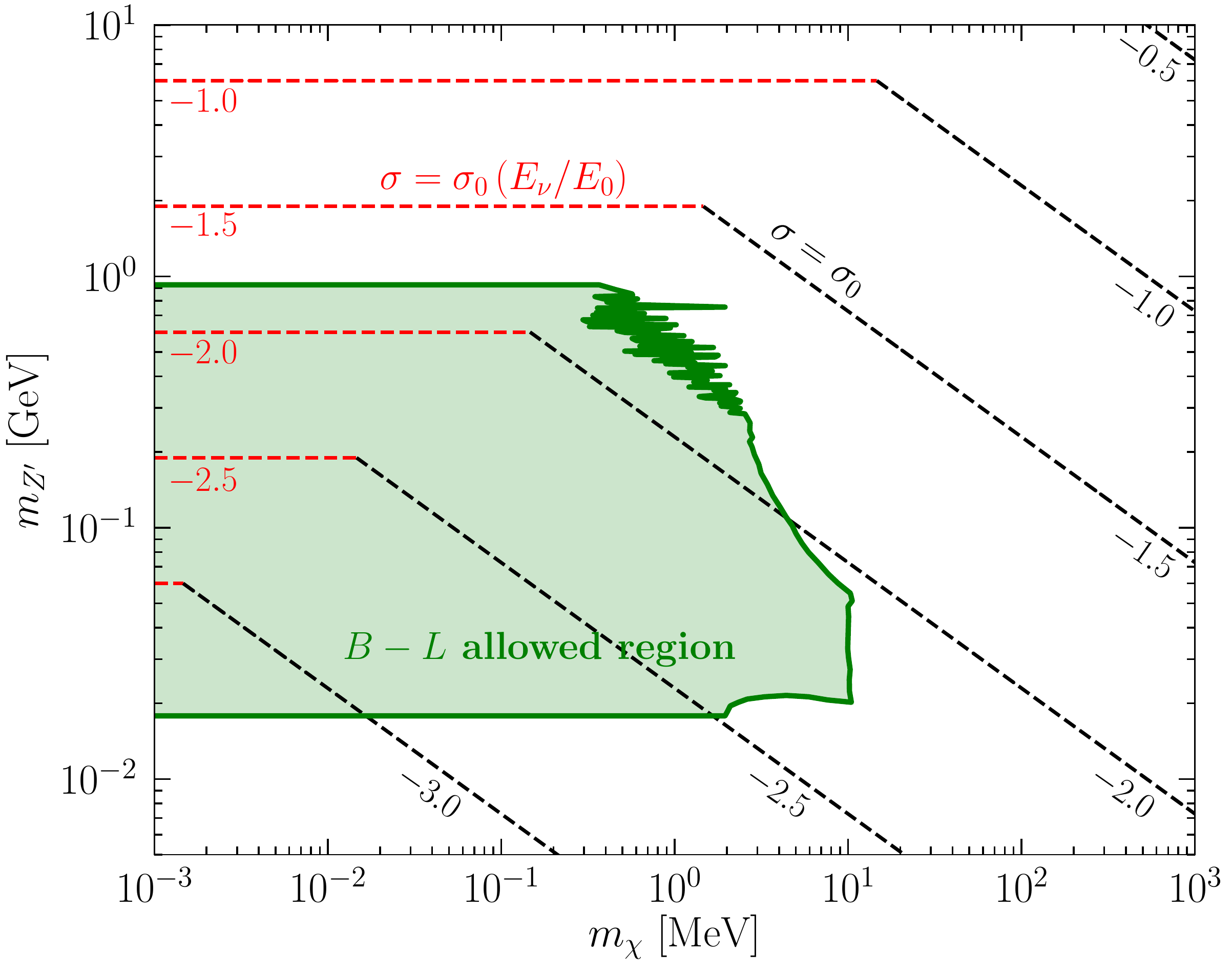}
\includegraphics[scale=0.34]{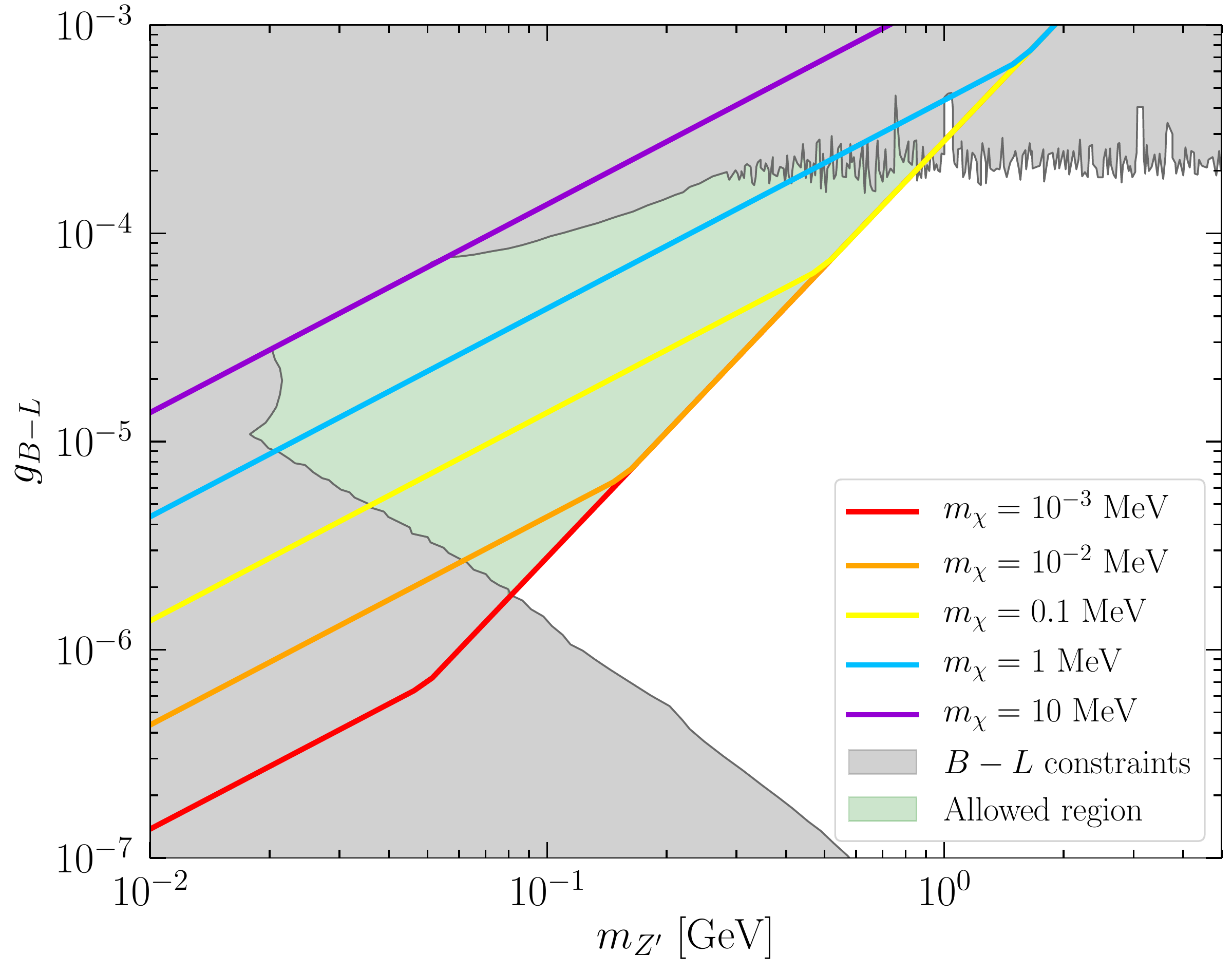}}
    \caption{\textit{Left:} Dashed lines are contours of $\log_{10}(\sqrt{g_\nu g_\chi}) \equiv \log_{10}g_{\rm eff}$ corresponding to the limit on the cross section from dark matter spike model BM1.
    Horizontal red (diagonal black) lines indicate where the cross section is approximately linear in (constant with) $E_\nu$.  Green region is that allowed for $U(1)'$ corresponding to $B-L$
    gauge symmetry.
    \textit{Right:} Constraints in the plane of
    $g_{B-L} = g_\nu$ versus $m_Z'$.  Gray regions are excluded by laboratory and astrophysical probes taken from Ref.~\cite{Bauer:2018onh}.  Colored lines are contours of constant dark matter mass such that the NGC 1068 constraint is saturated, for the BM1 DM spike model, with $g_\chi = 1$. The green shaded region corresponds to the same as in the left panel.
    } 
    \label{fig:plot}
\end{figure*}

\section{$Z'$ interpretation}
\label{sec:Zp_model}
To relate the derived constraints to an underlying particle physics model, we consider a new gauge boson $Z'$ that couples 
equally to all three flavors of leptons with strength $g_\nu$.  It is not sufficient
for our purpose to couple to a single flavor, because the new interaction induces a neutrino self-energy in the dark matter spike
\cite{Choi:2019zxy},  analogous to the Wolfenstein potential for electron neutrinos in the matter.  We find that for the cross sections of interest, it suppresses oscillations, such that the other two flavors would escape from the DM spike unimpeded.
A simple anomaly-free choice is to couple $Z'$ to baryon minus lepton number, $B-L$.

 For definiteness, we take the dark matter $\chi$ to be a complex scalar, with $B-L$ charge $g_\chi$, while the leptons have charge $g_\nu$.  The cross section for neutrinos of energy $E_\nu$ to scatter on DM at rest has the limiting behaviors~\cite{Cline:2022qld}
\be
\sigma ={g_\chi^2 g_\nu^2\over 4\pi m_{Z'}^2}\left\{\begin{array}{cc} 1,& E_\nu\gg {m_{Z'}^2/m_\chi}_{\phantom{|}}\\
    m_\chi E_\nu/m_{Z'}^2,& E_\nu\ll m_{Z'}^2/m_\chi
    \end{array}\right.\,.
    \label{sigma-limits}
\ee
This demonstrates the relevance of the choices of $\sigma$ being constant or linearly rising with $E_\nu$.

The left panel of Fig.\ \ref{fig:plot} illustrates the typical values of $m_{Z'}$
and the effective coupling $g_{\rm eff} = \sqrt{g_\chi g_\nu}$ that would saturate the 
constraint in Fig.\ \ref{fig:limit1} arising from the BM1 spike model.  
The
dashed lines indicate contours of $\log_{10}g_{\rm eff}$ for given choices of $m_{Z'}$ and $m_\chi$.  The entire region shown is consistent with perturbative unitarity, for $m_{Z'}\lesssim 130\,$GeV.  However, $g_\nu$ is
constrained by numerous experimental tests of the $B-L$ model,
summarized for example in Ref.\ \cite{Bauer:2018onh}.  The allowed parameter space is
indicated in Fig.\ \ref{fig:plot} by the shaded green region, assuming that $g_\chi\lesssim 1$.
 To match to the weaker limits corresponding to a different spike model $X$, the values of the couplings would have to be rescaled as 
$g_{\rm eff} \to g_{\rm eff}\times(\sigma_X/\sigma_{BM1})^{1/4}$.  For example with the BM1$'$ model, $g_{\rm eff}$ increases by a factor of $\sim 3$,
hence $\log_{10}g_{\rm eff} \to (\log_{10}g_{\rm eff} + 0.5)$.

A complementary view of our new constraint, versus existing limits on $g_\nu$ versus $m_{Z'}$, is shown in the right panel of  Fig.\ \ref{fig:plot}.  The gray region is
excluded by laboratory and astrophysical considerations.  The colored curves are contours of constant $m_\chi$ that saturate the BM1 constraint derived above, for the case of $g_\chi = 1$.  For smaller values of $g_\chi$, the curves would move upward by the the factor $1/g_\chi$.  The break in the curves for $m_\chi \lesssim 0.1\,$MeV occurs at the transition between constant cross section and  $\sigma$ linearly rising with energy.  The shaded green region therefore corresponds to parameter space where our new constraint can be saturated while remaining consistent with other probes of the 
$B-L$ model.

\section{Dark matter relic density}

It is striking that the $\nu$-DM interaction may play a crucial role in determining the relic density of the dark matter, through the annihilation process $\nu\nu\to\chi\bar\chi$.  For $m_\chi\gtrsim 1\,$MeV, this could occur through  thermal freezeout,  or by having an initial asymmetry between $\chi$ and its antiparticle, giving asymmetric DM (ADM).  The thermally averaged annihilation cross section
(times relative velocity) should satisfy $\langle\sigma_a v\rangle\sim
3\times 10^{-26}\,$cm$^3/$s \cite{Steigman:2012nb} for symmetric DM,
and it should be greater for ADM, in order to suppress the symmetric
component.

For lighter DM with $m_\chi \lesssim 1\,$MeV, thermal production is not viable since it results in warm dark matter, which is observationally disfavored.  Instead, the annihilation process can slowly build up the DM density over time, while remaining out of thermal equilibrium.
This is the freeze-in mechanism \cite{Hall:2009bx}, which requires much smaller values of
$\langle\sigma_a v\rangle$.

Although the annihilation cross section is not identical to the elastic scattering cross section, they are closely related in typical particle physics models.  In the model with a massive
$Z'$ gauge boson, at low energies the effective interaction Lagrangian is
\be
    {\cal L} = {1\over\Lambda^2}(\chi^*\!\!\stackrel{\leftrightarrow}{\partial}_\mu\!\!\chi)(\bar\nu\gamma^\mu\nu)
    \label{effop}
\ee
in the case of complex scalar DM, 
where $\Lambda = m_{Z'} / \sqrt{g_\nu g_\chi}$, the ratio of the $Z'$ mass to its couplings.  Then the elastic cross section scales as
$\sigma \sim {m_\chi E_{\nu}/\Lambda^4}$, while the annihilation cross section behaves as
\be
        \langle\sigma_a v\rangle \sim \left\{
        \begin{array}{lc} 
        m_\chi^2/\Lambda^4,& T\ll {m_\chi}_{\phantom{|}}\\
        {T^2/\Lambda^4}^{\phantom{|}},& m_{Z'} > T \gg {m_\chi}\\
        (g_\nu g_\chi)^2 /T^2,& T\gg m_{Z'}
        \end{array}\right.
        \label{sava}
\ee
The low-$T$ limit is appropriate for the case of freeze-out, while the
high-$T$ limit dominates in the case of freeze-in.   The third line of Eq.\ (\ref{sava})
indicates the behavior ensuing when the effective description in Eq.\ (\ref{effop}) breaks down due to the $Z'$ propagator.

\begin{figure}[t]
\centerline{\includegraphics[scale=0.34]{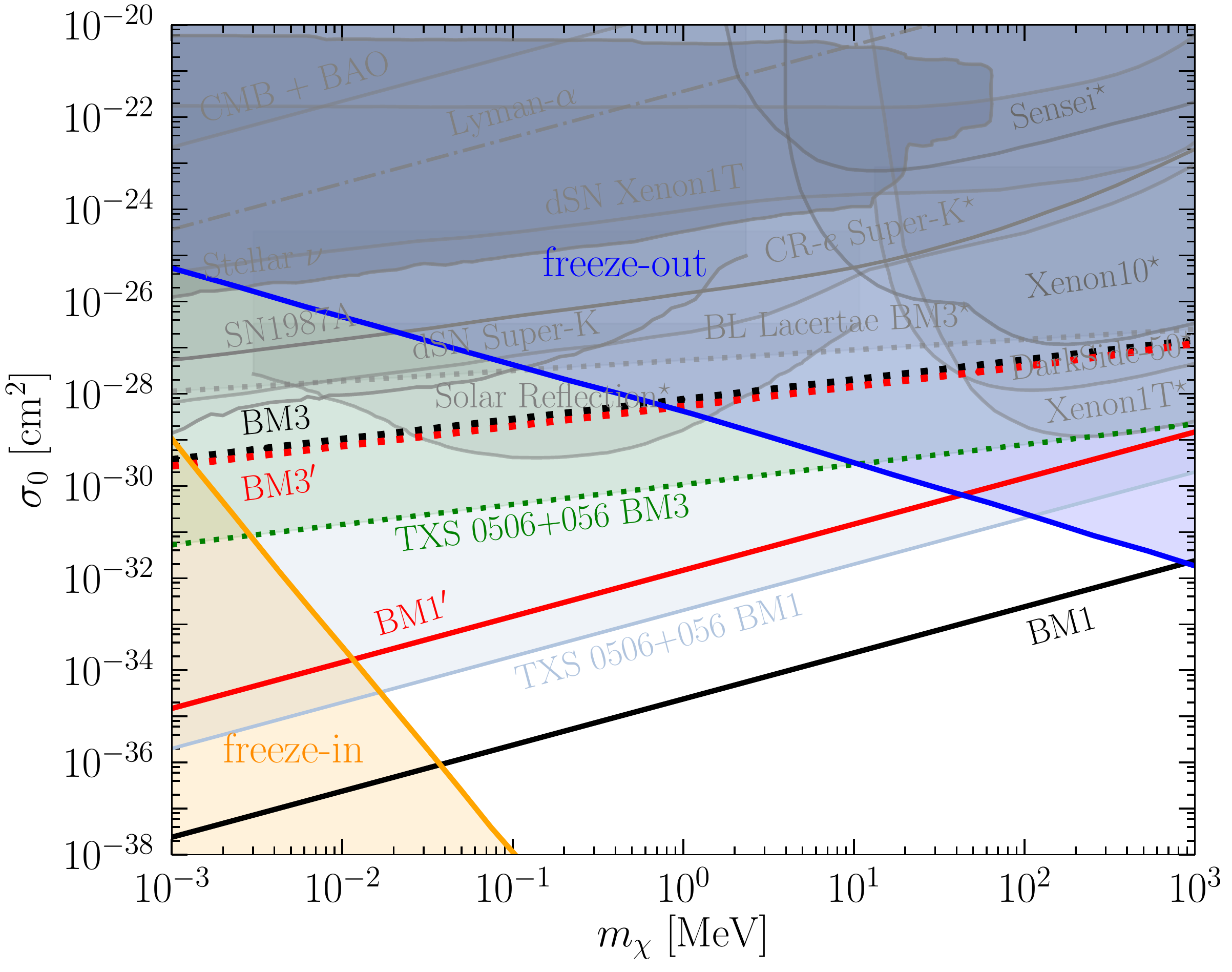}
    }  
    \caption{$90\%$ C.L. upper limits on the $\nu$-DM and $e$-DM scattering cross sections at the reference energy $E_0 = 10$ TeV, assuming they scale as $\sigma_{\ell \chi} \propto E_{\nu}$ with $\ell = \nu, e$ as shown in the right panel of Fig.~\ref{fig:limit1}. The regions of parameter space where freeze-out (blue curve) and freeze-in (orange curve) mechanisms can lead to the correct DM relic density are also shown. In particular, models for which $(m_{\chi}, \sigma_0)$ is above the blue or below the orange curves can accommodate only a sub-dominant faction of DM in the case of fully symmetric DM or leads to a suppression of the symmetric component in the case of ADM.
    } 
    \label{fig:fio}
\end{figure}

The freeze-in production is determined by the Boltzmann equation
(see for example Ref.\ \cite{Cline:2018fuq})
\be
    {dY_\chi\over dx} = {x\,s\,\langle\sigma_a v\rangle\over H(m_\chi)}
        Y_{eq}^2\,,
        \label{boltzeq}
\ee
where $Y_\chi$ is the DM abundance, $x=m_\chi/T$, $s$ is the entropy density, $H(m_\chi)$ is the Hubble parameter at $T=m_\chi$, and $Y_{eq}$ is the $\chi$ equilibrium density.  DM production is dominated by the high-$T$ contributions near $T\sim m_{Z'}$.  Integrating Eq.\ (\ref{boltzeq}) and imposing the observed relic density $Y_\chi m_\chi \cong 4\times 10^{-10}\,$GeV~\cite{Planck:2018vyg}, we find that
\be
    \left(\Lambda\over m_\chi\right)^4 \sim 5\times 10^6\left(\ln(m_{Z'}/m_\chi)\over g_{*}^{3/2}\right)^{1/4}\,,
\ee
which can be used to estimate the magnitude of $\sigma_0$ for the
case of $\nu$-DM scattering with linear energy dependence.  

This procedure results in the heavy orange line (labeled ``freeze-in'') on Fig.\ \ref{fig:fio}.  Values above the line would lead to 
overabundant dark matter, in the absence of an additional annhilation channel.  Similarly, one can estimate that $m_\chi/\Lambda^2\sim 5\times 10^{-5}$\,GeV$^{-1}$ for the desired relic abundance through thermal freeze-out.  This determines $\sigma_0$
versus $m_\chi$ as shown by the heavy blue line (labeled ``freeze-out'')
in Fig.\ \ref{fig:fio}.
Values of $\sigma_0$ above the line correspond to a subdominant DM component, in the case of fully symmetric DM, or suppression of the symmetric component in the case of ADM.  Dark matter in the region of parameter space between the two lines would require some additional interactions for achieving the observed relic density.

Comparing to the results of the previous section, we see that the freeze-out scenario is not viable for pure $B-L$
gauge models that saturate the BM1 constraint, but freeze-in is consistent.

\section{Conclusions}

IceCube has now observed neutrinos from 
two active galactic nuclei, and other researchers have suggested that additional
AGNs have produced excess IceCube events with lower statistical significance~\cite{Rodrigues:2020fbu,Giommi:2020viy,Fermi-LAT:2019hte,Kadler:2016ygj,Sahakyan:2022nbz,IceCube:2021slf,Giommi:2020hbx,Franckowiak:2020qrq}.  It seems likely that in the coming years, more such sources will be discovered, which may further strengthen 
constraints on dark matter-neutrino interactions.

In the present work, we derived new limits on $\nu$-DM scattering, which depend strongly on the details of the DM spike density surrounding the AGN's central supermassive black hole, in  particular, the power $\alpha$ in the
assumed density profile $\rho\sim r^{-\alpha}$.  Previous work on the blazar TXS 0506+056 considered only the case $\alpha = 7/3$ \cite{Ferrer:2022kei,Choi:2019ixb}, which
neglects relaxation of the 
initial DM spike by gravitational scattering with stars, whereas
$\alpha \to 3/2$ when this effect is maximal.
Here we have considered both possibilities, as did by Ref.~\cite{Bhowmick:2022zkj} in the context of dark matter boosted by electrons in the blazar jet.
Probably $\alpha=3/2$ is more realistic due to the complex stellar environment in AGNs~\cite{Romeo:2016hms}, but we do not feel qualified to
make a definitive statement, and we defer to workers who specialize on this issue to confirm such a statement, which would significantly reduce the astrophysical uncertainty of the constraints.

Other astrophysical uncertainties that can significantly affect the predictions for AGNs 
concern the position $R_{\rm em}$ within the relativistic jet where neutrinos are likely to be produced, and the choice of the initial neutrino flux $\Phi_{\nu}$.  In these respects,
radio galaxies like NGC 1068 allow for more robust predictions than blazars.
Observations of different radio galaxies (see e.g.~\cite{Gallimore:2004wk,Inoue:2018kbv}) suggest that the black hole corona, where neutrinos with energy of $\sim \mathcal{O}(10\,\,\text{TeV})$ can be generated, has to extend at most several tens of the Schwarzschild radii $R_S$ from the galactic center~\cite{Murase:2022dog,Halzen}, which is one to two orders of magnitude smaller than the radius of the neutrino-emitting region in blazar jets~\cite{Padovani:2019xcv}.  
Furthermore, the initial neutrino flux we used to set constraints on the $\nu$-DM scattering comes directly from the IceCube observation of NGC 1068 and does not rely on the result of numerical simulations, which might be affected by simplifying assumptions in modelling the astrophysical source. 
Choosing a different initial neutrino spectrum, as derived from theoretical modeling or simulations of NGC 1068, generally strengthens our results or, in the worst case scenario, weakens the limits by a factor of a few.

The remaining important uncertainty is on whether
dark matter can annihilate within the DM spike,
and this cannot be easily resolved.  However we can say that a consistent picture emerges if the DM obtains its relic density through freeze-in, mediated by the same interactions that we are constraining.  In this case the DM annihilation cross section can be much smaller than the values where annihilation would significantly deplete the spike, and our strongest limits would then apply.

\bigskip
{\bf Acknowledgment.}  We thank Andrew Benson, Daryl Haggard and Shengqi Yang for helpful information about NGC 1068 structure, Paolo Gondolo about the profile of the dark matter spike, Kohta Murase for clarification on the neutrino emitting region and astrophysical uncertainties, Francis Halzen for an alternative interpretation of the IceCube neutrino event from TXS 0506+056,
and Matthew Lundy and Samantha Wong  for stimulating discussion. We also thank Ken Ragan, Aaron Vincent and Andreas Warburton for clarifications and suggestions that improved the original manuscript.  This research was supported by the Natural Sciences and Engineering Research Council (NSERC) of Canada.

\bibliographystyle{utphys}
\bibliography{ref2.bib}

\end{document}